\documentstyle[12pt]{article}

\begin{document}
\newcommand{\od}{\stackrel{\cdot}}
\newcommand{\td}{\stackrel{\cdot\cdot}}
\newcommand{\fd}{\stackrel{\cdot\cdot\cdot}}

\topmargin 0pt
\oddsidemargin 5mm
\begin{titlepage}
\setcounter{page}{0}

\vspace{2cm}
\begin{center}
{\Large Localization of phonons in two component
superlattice with random thicknesses
of the layers}
\vspace{1cm}

{\large David G. Sedrakyan} \\
{\em Yerevan State University, Armenia}\\
\vspace{1cm}
{\large Ara G. Sedrakyan \footnote{e-mail: {\sl sedrak@lx2.yerphi.am ;
sedrak@nbivax.nbi.dk}}}\\
{\em Yerevan Physics Institute, Armenia}
\end{center}
\vspace{5mm}
\centerline{{\bf{Abstract}}}

Two component superlattice film of $2N$ layers is
considered and dimensionally quantised spectrum of
phonon-s is found.

The problem of localization of phonon-s in the superlattice
with random thicknesses of the layers is investigated.
The Landauer resistance of the transport of phonon-s
is calculated exactly. For short range disorder the 
numerical analyses shows, that
at frequency $\omega$=0, there is delocalized state
and the correlation length index $\nu$ is equal to 2.

\end{titlepage}
\newpage
\renewcommand{\thefootnote}{\arabic{footnote}}
\setcounter{footnote}{0}

\section{Introduction}
\indent
The interest to superlattice structure in condensed matter
physics is based on the possibility of manipulation of the
physical properties of devices by changing of characteristic
lattice parameters.

The growth techniques can be used to prepare specimens consisting
of alternating layers of thickness $d_1$ of constituent $A$
and thickness $d_2$ of constituent $B$. Samples can be prepared
so that $d_1$ and $d_2$ have any value from two or three atomic
spacings  up to hundred   atomic spacings. The entities $A$ and $B$
can be materials with different acoustic, electronic, magnetic 
properties and one can consider semiconductor, metal, insulator 
and superconductor
constituents, attempting to change values of expectable physical 
variables in a desirable regime.

The technical advance in fabrication of superlattices motivated 
an intensive  study of various physical
properties of these systems, especially electronic and vibration 
spectra, optical and magnetic properties, e.t.c. The calculations
\cite{2,3} shows that the spectrum 
of quasi-periodic systems are intermediate between
periodic and random ones.

Along with electronic properties study of elastic waves in a bulk
superlattices has been a subject of interest in the last decades
\cite{8,9,10,11,12,13,14,15,16,17,18,19}.
A one dimensional theory of acoustic vibrations in layered
material was given long ago by Rytov \cite{1}. Elastic waves
have also been investigated in semi-infinite superlattices
\cite{20,21,22,23,23}.

The consideration of the superlattice films instead of the
massive ones provides additional opportunities for controlling
the elastic and electronic parameters of the superlattices. A
sufficiently complete experimental knowledge about the
oscillator spectra of binary laminated  semiconductor systems is
available nowadays \cite{24}. Numerous and generally mutually
compatible results about the frequencies of long-wave phonon-s in
$In As-Ga Sb, Ge-Ga As$ are presented in literature \cite{25}.

Together with strongly periodic superlattices the effects of
localization and tunnelling of the electrons was studied in
short rang disorder superlattices \cite{4,5,6,7}

The problem of localization of electrons in random potential 
and hopping parameters in low dimensional spaces are in continuous 
interest
of physicists after Andersons remarkable article \cite{An}.
Originally Mott and Twose \cite{MT} conjectured that all states 
are localized in $1D$ systems for any degree of disorder.
It was argued \cite{Ab} that in case of full randomness all
states are localized in dimensions equal and less of two.
However, recent investigations shows \cite{DIS1,DIS2,DIS3}
that delocalized states can appear in case of correlated disorder.

The aim of present article is twofold. First we found the spectrum
of transversal phonon-s  in two component superlattice film with 
boundaries with arbitrary finite number of slices. 
The results, obtained here, are also applicable
 for longitudinal waves, which moves in the transversal to the layers
 direction. Second, we consider random distribution of the
thicknesses of the superlattice compounds and calculate the
Landauer resistance \cite{La} of the acoustic phonon-s exactly.
The continuous model is used, in which the layers are
 considered as a macroscopic elastic bodies.

\section{Notations and equation of motion in the bulk}
 \indent

   We follow here the notations and derivations of the book \cite{26}.

Let the vector of elastic displacement (deformation) of the 
material at
the space point $\vec{x}$ is $\vec{u}(\vec{x})$. For small
variations $(\vec{u}<<1)$ we can derive the strain tensor (tensor of
deformation) as
\begin{equation}
\label{A1}
u_{ik}=\frac{1}{2}\left(\frac{\partial u_i}{\partial x^k}+
\frac{\partial u_k}{\partial x^i}\right)\quad \quad i,k=1,2,3.
\end{equation}
The response of the free energy $F$ to variation of deformation
tensor defines the stress tensor $\sigma_{ik}$ as
\begin{equation}
\label{A2}
dF=-SdT+\sigma_{ik}du_{ik},
\end{equation}
According to Hooke's law, for small deformations  $u_{ik}$
around equilibrium, we can restrict ourself to consider only
quadratic in  $u_{ik}$ terms in series expansion of $F$. Define
\begin{equation}
\label{A3}
dF=\left[ Ku_{ll}\sigma_{ik}+2\mu\left( u_{ik}-\frac{1}{3}
u_{ll}\delta_{ik}\right)\right]d u_{ik},
\end{equation}
which defines the following stress tensor
\begin{equation}
\label{A4}
\sigma_{ik}=Ku_{ll}\sigma_{ik}+2\mu\left( u_{ik}-\frac{1}{3}
\delta_{ik}u_{ll}\right),
\end{equation}
where $K$ is the modulus of compression and $\mu$ is the modulus of
rigidity. $K>0$ and $\mu>0$. One can express stress $\sigma_{ik}$
in  terms of Young's modulus $E=\frac{3k\mu}{3k+\mu}$ and Poisson's
coefficient $\sigma=\frac{1}{2}\,\frac{3k-3\mu}{3k+\mu}$ as
follows
\begin{equation}
\label{A5}
\sigma_{ik}=\frac{E}{1+\sigma}\left(
u_{ik}+\frac{\sigma}{1-2\sigma}
u_{ll}\delta_{ik}\right).
\end{equation}
Then, the equation of motion for the displacement $u_i$ is simply
\begin{equation}
\label{A6}
\rho\td{u}_i=\frac{\partial\sigma_{ik}}{\partial x_k}
\end{equation}
where $F_i=\frac{\partial\sigma_{ik}}{\partial x_k}$ is the 
force, which acts on unite volume around $\vec{x}$ with density of
matter $\rho(\vec{x})$.  After substituting (\ref{A4}) into
(\ref{A6}) we can define the equation for $u_i(x)$ in a bulk.
\begin{equation}
\label{A7}
\rho \td{u}_i=\frac{E}{2(1+\sigma)} \Delta u_i + 
\frac{E}{2(1+\sigma)
(1-2\sigma)} {\rm grad} \frac{\partial u_i}{\partial x^i}.
\end{equation}
For the transversal waves
\begin{equation}
\label{A8}
{\rm div}\vec{u}=0,
\end{equation}
and from the (\ref{A7}) we can easily get
\begin{equation}
\label{A9}
\frac{\partial^2 u_i}{\partial t^2}-c_t^2 \Delta u_i=0,
\end{equation}
with
\begin{equation}
\label{A10}
c_t^2=\frac{E}{2\rho(1+\sigma)}.
\end{equation}
For the longitudinal waves
\begin{equation}
\label{A11}
\Delta \vec {u} = {\rm grad} 
({\rm div}\vec{u})
\end{equation}
the equation (\ref{A7}) reduces to the following equation
\begin{equation}
\label{A12}
\frac{\partial^2 u_i}{\partial t^2}-c_e^2\, \Delta u_i=0
\end{equation}
where $c_{l}$ is the longitudinal velocity
\begin{equation}
\label{A13}
c_{l}^2 =\frac{E(1-\sigma}{\rho(1+\sigma)(1-2\sigma)}.
\end{equation}
\section{Boundary conditions in the superlattice film, Transfer
Matrix and the spectrum}
 \indent

Let's consider superlattice, elementary cells of which consists
 of two layers of various materials  $A$ and $B$ with the
thickness $d_1, d_2$ and  modulus of rigidity $\mu_1, \mu_2$ (fig.1).  
The number of pairs in the film is $N$.

\vspace{2cm}
\setlength{\unitlength}{0.7cm}
\hspace{2cm}
\begin{picture}(15,7)
\put(1,4){\vector(1,0){14.5}}
\put(1,4){\vector(0,1){3}}
\put(1,4){\vector(-1,-1){1.5}}
\put(2,3){\line(0,1){4}}
\put(2.4,6.2){\shortstack{$d_1$}}
\put(2.4,4.5){\shortstack{A}}
\put(2.5,6){\vector(-1,0){.5}}
\put(2.5,6){\vector(1,0){.5}}
\put(3,3){\line(0,1){4}}
\put(3.8,6.2){\shortstack{$d_2$}}
\put(3.8,4.5){\shortstack{B}}
\put(4,6){\vector(-1,0){1}}
\put(4,6){\vector(1,0){1}}
\put(5,3){\line(0,1){4}}
\put(6,3){\line(0,1){4}}
\put(8,3){\line(0,1){4}}
\put(11,3){\line(0,1){4}}
\put(12,3){\line(0,1){4}}
\put(14,3){\line(0,1){4}}
\multiput(8.2,6)(0.5,0){6}{\circle*{.1}}
\put(0.5,6.5){\shortstack{y}}
\put(0,2.5){\shortstack{x}}
\put(15,3.5){\shortstack{z}}
\put(11.4,4.5){\shortstack{A}}
\put(12.8,4.5){\shortstack{B}}
\put(6.5,1){\shortstack{Fig.1.}}
\end{picture}

We  consider transversal elastic waves $({\rm div } \vec{u}=0)$
propagating inside superlattice in arbitrary direction. It can be
shown, that all results are reproducible for longitudinal waves if
they propagates in transversal to layers direction.

Lets choose coordinate system such, that waves are propagating in
the $(x,y)$ plane ($x$ represents transversal to layers direction), 
with the wave vectors $(k_1, q_1, 0)$ and
$(k_2, q_2, 0)$ in the $A$ and $B$ materials correspondingly.
Without loose of generality one can take $u_x=u_y=0$.

The solutions of the wave equations (\ref{A9}) with frequency $\omega$,
which full-fills transversality condition ${\rm div}\vec{u}=0$,
is the superposition of forward - and a backward-travelling
waves
\begin{eqnarray}
\label{A14}
&u_{2n-1}=\left(c_{2n-1}e^{ik_1x}+\bar{c}_{2n-1}e^{-ik_1x}
\right)e^{i(q y-wt)},\nonumber\\
&u_{2n}=\left(c_{2n}e^{ik_2x}+\bar{c}_{2n}e^{-ik_2x}
\right)e^{i(q y-wt)}, \;\;\;\;\;\;\;\;\;  &n=1\cdots N, 
\end{eqnarray}

where for $k_i,  i=1,2$ we have
\begin{equation}
\label{A15}
k_i^2+q^2=\frac{\omega^2}{c_{it}^2}
\end{equation}

In formulas  (\ref{A14})  $2n$(correspondingly$2n-1$) numerates 
the layers of $B$(or $A$)
type and $ c_{1t}(c_{2t})$ are the velocities of sound in that
materials.

We should now impose the
boundary conditions on the displacements $u_{2n}$ and $u_{2n-1}$.

Let's  consider now free boundaries of the film,  which means the  
use of
Neumann boundary conditions

\begin{equation}
\label{A16}
\partial_x u_1=\partial_x u_{2N}=0.
\end{equation}

On the boundary of the   $A$ and $B$  layers one should use the
continuity condition for the displacements 
\begin{equation}
\label{A17}
u_{2n}=u_{2n-1},
\end{equation}
as well for the forces
\begin{equation}
\label{A18}
F_i = \sigma_{ik}ds_k.
\end{equation}
 In the equation (\ref{A18}) $ds_k = n_k ds$-is the normal 
vector to the boundary and equal to small
area in modulo.
Hence we have
\begin{equation}
\label{A19}
\sigma^1_{ik}n_k=\sigma^2_{ik}n_k.
\end{equation}
On the boundaries of the film the forces are equal to zero
\begin{equation}
\label{A20}
\sigma_{ik}n_k=0
\end{equation}
By use of expression (\ref{A5}) for $\sigma_{ik}$ from the
boundary conditions (\ref{A16})-\ref{A17})  and
(\ref{A19}-\ref{A20}) one can easily obtain following   set of
equations for the displacements $u$ 
\begin{eqnarray}
\label{A21}
& u_x^{2N}=u^1_x=0,\nonumber\\
& \mu_2 u^{2n}_x=\mu_1 u^{2n-1}_x,\nonumber\\
& u^{2n}=u^{2n-1}& n=1, 2, \cdots N.
\end{eqnarray}
This equations transforms into the following equations for  
the coefficients of
the forward and backward travelling waves
\begin{eqnarray}
\label{A22}
&c_{2N}e^{ik_2(d_1+d_2)n}-\bar{c}_{2N}e^{-ik_2(d_1+d_2)n}= 0
\nonumber\\
&\mu_2 k_2 c_{2n}e^{ik_2[(d_1+d_2)(n-1)+d_1]}-\mu_2k_2\bar{c}_{2n}
e^{-ik_2[(d_1n+d_2(n-1)]}=\nonumber\\
&=\mu_1 k_1 c_{2n-1}e^{ik_1 [d_1n+d_2(n-1)]}-\mu_1
k_1\bar{c}_{2n-1}e^{-ik_1[d_1n+d_2(n-1)]}\nonumber\\
&c_{2n}e^{ik_2[d_1n+d_2(n-1)]}+\bar{c}_{2n}
e^{-ik_2[d_1n+d_2(n-1)]}=\nonumber\\
&=c_{2n-1}e^{ik_1[d_1n+d_2(n-1)]}+\bar{c}_{2n-1}
e^{-ik_1[d_1n+d_2(n-1)]}\\
&\vdots\nonumber\\
&c_1-\bar{c}_1= 0\nonumber
\end{eqnarray}
We will solve this set of linear equations  by use of
transfer matrix method \cite{8}.

Let's define now
\begin{equation}
\label{A23}
\psi_{2n}=\left(\begin{array}{l}
c_{2n}\\
\bar{c_{2n}}
\end{array}\right).
\end{equation}
Then the half of the set of equations (\ref{A23})can be 
reformulated as
follows
\begin{equation}
\label{A24}
A_{2n}\psi_{2n}=B_{2n-1} \psi_{2n-1},
\end{equation}
with
\begin{equation}
\label{A25}
A_{2n}=\left( \begin{array}{ll}
e^{ik_2\left( nd_1+(n-1)d_2\right)},& -e^
{-ik_2\left( nd_1+(n-1)d_2\right)}\\
e^{ik_2\left( nd_1+(n-1)d_2\right)}, & e^{-ik_2\left(
nd_1+(n-1)d_2\right)}
\end{array}\right),
\end{equation}
and
\begin{equation}
\label{A26}
B_{2n-1}=\left( \begin{array}{ll}
{\mu_1k_1 \over \mu_2k_2}e^{ik_1\left( nd_1+(n-1)d_2\right)},
& -{\mu_1k_1 \over \mu_2k_2}e^{-ik_1\left( nd_1+(n-1)d_2\right)}\\
e^{ik_1\left( nd_1+(n-1)d_2\right)}, & e^{-ik_1\left(
d_1n+(n-1)d_2\right)}
\end{array}\right).
\end{equation}
Equation (\ref{A24}) can be rewritten as
\begin{equation}
\label{A27}
\psi_{2n}=A_{2n}^{-1}\, B_{2n-1}\, \psi_{2n-1}.
\end{equation}
Similarly, the other half of the equations (\ref{A22}) 
looks as
\begin{equation}
\label{A28}
\psi_{2n-1}=A_{2n-1}^{-1}\, B_{2n-2}\, \psi_{2n-2},
\end{equation}
with
\begin{equation}
\label{A29}
A_{2n-1}=\left( \begin{array}{ll}
e^{ik_1\left( (n-1)d_1+(n-1)d_2\right)},&
-e^{-ik_1\left((n-1)d_1+(n-1)d_2\right)}\\
e^{ik_1\left((n-1)d_1+(n-1)d_2\right)}, & e^{-ik_1\left(
(n-1)d_1+(n-1)d_2\right)}
\end{array}\right),
\end{equation}
and
\begin{equation}
\label{A30}
B_{2n-2}=\left( \begin{array}{ll}
{\mu_2k_2 \over \mu_1k_1}e^{ik_2\left[(n-1)d_1+(n-1)d_2\right]}, 
& -{\mu_2k_2 \over \mu_1k_1}e^{-ik_2\left[(n-1)d_1+(n-1)d_2\right]}\\
e^{ik_2\left[(n-1)d_1+(n-1)d_2\right]}, & e^{-ik_2\left[
(n-1)d_1+(n-1)d_2\right]}
\end{array}\right).
\end{equation}
Recursion equations (\ref{A27}) and (\ref{A28}) allows us to
connect
$\psi_{2n}$ with $\psi_1$ in a following way
\begin{equation}
\label{A31}
\psi_{2n}=A_{2n}^{-1}\, B_{2n-1}\,A_{2n-1}^{-1}\,
B_{2n-2}\cdots
A_2^{-1} \, B_1\, \psi_1.
\end{equation}
Let's now define the Transfer Matrices as
\begin{equation}
\label{A32}
T_1=B_{2n-1}\,A_{2n-1}^{-1}=\left( \begin{array}{ll}
{\mu_1k_1 \over \mu_2k_2}\cos k_1d_1 &
i{\mu_1k_1 \over \mu_2k_2}\sin k_1d_1\\
i \sin k_1d_1 & \cos k_1 d_1
\end{array}\right)
\end{equation}

and
\begin{equation}
\label{A33}
T_2=B_{2n}\,A_{2n}^{-1}=\left( \begin{array}{ll}
{\mu_2k_2 \over \mu_1k_1}\cos k_2d_2 &
i{\mu_2k_2 \over \mu_1k_1}\sin k_2d_2\\
i \sin k_2d_2 & \cos k_2 d_2
\end{array}\right).
\end{equation}
Then the equation (\ref{A31}) becomes
\begin{equation}
\label{A34}
\psi_{2n}=B_{2n}^{-1}\, T^{n}\,A_1\,\psi_1 =U \psi_1
\end{equation}
where
\begin{equation}
\label{A35}
T=T_1 T_2=\left( \begin{array}{ll}
\cos k_1d_1 \cos k_2 d_2 -
&
i\cos k_1d_1\sin k_2d_2 +\\
- \frac{\mu_1k_1}{\mu_2k_2}\sin k_1d_1\sin k_2 d_2,
& + \frac{i\mu_1k_1}{\mu_2k_2}\sin k_1d_1 \cos k_2d_2\\
&\\
i\sin k_1d_1\cos k_2 d_2 + &
- \frac{\mu_2k_2}{\mu_1k_1}\sin k_1d_1\sin k_2d_2 +\\
+ \frac{i \mu_1k_1}{\mu_2k_2}\cos k_1d_1\sin k_2d_2
&+ \cos k_1d_1 \cos k_2 d_2
\end{array}\right)
\end{equation}
The first equation of (\ref{A22}) for $n=N$ can be 
written as
\begin{equation}
\label{A36}
\bar{\psi}_{2N} \psi_{2N}=0,
\end{equation}
where
\begin{equation}
\label{A37}
\bar{\psi}_{2N}=\left(e^{ik_2\left(N d_1+N d_2\right)}, 
\quad -e^{-ik_2
\left(N d_1+N d_2\right)}\right).
\end{equation}
For the another boundary of the superlattice film, where
 $n=1$ we have
\begin{equation}
\label{A38}
c_1=\bar{c}_1=c,
\end{equation}
which means that
\begin{equation}
\label{A39}
\psi_1=c \left( \begin{array}{l}
1\\
1
\end{array}\right).
\end{equation}
From the equations (\ref{A34}) and (\ref{A36}) we can obtain
following equation for the spectrum
of transversal phonon-s in the superlattice film of $2N$ layers
\begin{equation}
\label{A40}
Tr\left[C T^{N}\right]=0,
\end{equation}
where
\begin{eqnarray}
\label{A41}
C^\alpha_\beta=(A_1\psi_1)_\beta(\bar{\psi}_{2N}B^{-1}_{2N})^\alpha
=\left( \begin{array}{ll}
0 & 0\\
2\frac{\mu_1k_1}{\mu_2k_2} & 0 \end{array}\right).
\end{eqnarray}

To proceed further we need to calculate the $N-1$ degree
of the  Transfer Matrix, which can be achieved simply
by diagonalizing $T$. Obviously
\begin{equation}
\label{A43}
T^{N} = W^{-1} \left(\begin{array}{ll}
\lambda^{N},\,\,0\\
0,\,\,\bar{\lambda}^{N}
\end{array}\right) W,
\end{equation}
where $\lambda$ and $\bar{\lambda}$ are eigenvalues of $T$, 
and $W$
is the diagonalizing matrix. 
One easily can find the eigenvalues or the Transfer Matrix $T$ as
\begin{eqnarray}
\label{A45}
\lambda=e^{\pm i\Theta}&=&\left(\cos k_1d_1 \cos k_2d_2-\frac{1}{2}
\left( \frac{\mu_1k_1}{\mu_2k_2}+
 \frac{\mu_2k_2}{\mu_1k_1}\right)
\sin k_1d_1\sin k_2d_2\right) \pm \nonumber\\
&\pm& i\sqrt{1-\left(\cos k_1d_1 \cos k_2d_2 -\frac{1}{2}
\left( \frac{\mu_1k_1}{\mu_2k_2}+
 \frac{\mu_2k_2}{\mu_1k_1}\right)
\sin k_1d_1\sin k_2d_2 \right)^2},
\end{eqnarray}
where
\begin{equation}
\label{A46}
\cos \Theta=\cos k_1d_1 \cos k_2d_2-\left( \frac{\mu_1k_1}{\mu_2k_2}+
 \frac{\mu_2k_2}{\mu_1k_1}\right)
\sin k_1d_1\sin k_2d_2.
\end{equation}

Further, a simple calculations shows, that equation (\ref{A40}) 
reduces to
\begin{equation}
\label{A48}
Im{\lambda^N}=0,
\end{equation}
which means that
\begin{equation}
\label{A49}
\Theta=\pi\frac{Q}{N},\;\;\;Q=1 \cdots N.
\end{equation}
Finally we obtain  following equation for the spectrum of
transversal phonon-s
\begin{equation}
\label{A50}
\cos k_1d_1 \cos k_2d_2-\left( \frac{\mu_1k_1}{\mu_2k_2}+
 \frac{\mu_2k_2}{\mu_1k_1}\right)
\sin k_1d_1\sin k_2d_2=\pm \cos \pi\frac{Q}{N}
\end{equation}
where
\begin{equation}
\label{A51}
k_i^2=\frac{\omega^2}{c_i^2}-q^2
\end{equation}
We see that this equation is coinciding with the
equation for the spectrum
of phonon-s in the bulk \cite{1,8}, but the momentums in
perpendicular to the layers direction are quantised
due to dimensional restriction of the film.
\section{The Landauer resistance of phonon-s in the
superlattice with random distribution 
of thicknesses of the layers}
\indent

The problem of elastic waves in superlattice is essentially
one dimensional. One dimensional problems are especially
attractive because of their possible exact integrability.
In the article by Erdos and Herndon \cite{EH} the problem
of the transport of particles in the one dimensional
space for a wide class of disorders was considered in the 
Transfer Matrix approach and general results were obtained.
It was proved that that Transfer Matrix of the one dimensional
problem belongs to $SL(2,R)$ group and randomness can be
exactly taken into account for such quantities as Landauer
resistance \cite{La}.

Some exact results for Kronig-Penney model in case of non-diagonal
disorder by other methods was obtained in \cite{Gas}.

It is easy to see from the formulas (\ref{A35}) for the Transfer
Matrix $T$, here we also have a representative of the $SL(2,R)$
group. One can make a link between transfer Matrices of the
Kronig-Penney model and phonon-s in the superlattice.

Following \cite{La} and \cite{EH} let's define resistance
as a ratio of reflection to transmission coefficients,
which, by use of formula (\ref{A34}), is
equal to
\begin{equation}
\label{RO}
\rho=\frac{1-|\tau|^2}{|\tau|^2}=U_{12} U_{12}^{*}=U^1_2 (U^+)^2_1,
\end{equation}
where
$U^1_2$ is the 1,2 matrix element of the evolution matrix $U$
\begin{equation}
\label{U}
U= B^{-1}_{2N}(I_1 T_2)^N A_1.
\end{equation}

We are going to consider random distribution of thicknesses
of the layers and take the average of Landauer resistance.
For further convenience we will normalise $T_1(T_2)$ Transfer
Matrices on order to have a unit determinant. It will not
change the equation (\ref{U}) because the normalisation
factors for $T_1$ and $T_2$ cancels each other. Hence we
will consider
\begin{equation}
\label{D}
T_{2i}=\left( \begin{array}{ll}
({\mu_2k_2 \over \mu_1k_1})^{1/2}\cos k_2(x_{2i}-x_{2i-1}) &
i({\mu_2k_2 \over \mu_1k_1})^(1/2)\sin k_2(x_{2i}-x_{2i-1})\\
i({\mu_1k_1 \over \mu_2k_2})^{1/2} \sin k_2(x_{2i}-x_{2i-1}) & 
({\mu_1k_1 \over \mu_2k_2})^{1/2}\cos k_2(x_{2i}-x_{2i-1})
\end{array}\right)
\end{equation} 
for the even slices. The similar expression for odd slices
$T_{2i-1}$ can be found simply by permuting variables
$k$ and $\mu$ for 1 and 2.

Now let us analyse the direct product of the evolution matrices,
the $(U \otimes U^{+})^{1,2}_{2,1}$ matrix element of
which defines Landauer resistance. For this purpose we should 
calculate first the simplest constituent block of that
expression, namely direct product $T_i \otimes T^{+}_i$
of $U_i$-s. In the article \cite{EH} it was demonstrated,
that this direct product can be represented as 
$1 \oplus (3$x$3)$ =$4$ x $4$ matrix. It happened because
of the fact, that $T_i$ matrices are a spinor representations
of the $SL(2,R)$, hence, the direct product of two
$1/2$-representations can be expanded as a sum of scalar and 
vector representations. In the language of the group elements
$T \in SL(2,R)$ this expansion looks like 
\begin{equation}
\label{DD}
(T_i)^{\alpha}_{\alpha'}(T_{i}^{-1})^{\beta'}_{\alpha}=
{1 \over 2}\delta^{\alpha}_{\beta} \delta^{\beta'}_{\alpha'}-
{1 \over 2}(\sigma^{\mu})^{\beta'}_{\alpha'} 
\Lambda_{i}^{\mu\nu}(\sigma^{\nu})^{\alpha}_{\beta},
\end{equation}
where
\begin{equation}
\label{D1}
\Lambda_{i}^{\mu\nu}={1 \over 2} Tr(T_i \sigma^{\mu} T_{i}^{-1}
\sigma^{\nu})
\end{equation}
is the spin-one part of the direct product. But for
Landauer resistance we need to calculate $T \otimes T^+$.
It is easy to see from the formula (\ref{D}) that
\begin{equation}
\label{D2}
\sigma_1 T^{-1} \sigma_1 = T^{+},
\end{equation}
therefore, by multiplying the the expression (\ref{DD})
in the left and right by $\sigma_1$ we will have
\begin{equation}
\label{D3}
(T_i)^{\alpha}_{\alpha'}(T_{i}^{+})^{\beta'}_{\beta}=
{1 \over 2}(\sigma_1)^{\alpha}_{\beta} 
(\sigma_1)^{\beta'}_{\alpha'}-
{1 \over 2}(\sigma^{\mu} \sigma_1)^{\beta'}_{\alpha'} 
\Lambda_{i}^{\mu\nu}(\sigma^{\nu} \sigma_1)^{\alpha}_{\beta}.
\end{equation}

Now the calculation of the direct product $U \otimes U^{+}$
is straightforward. The product $\prod_{i=1}^{2N}T_i$ of
$T_i$-s transforms into product of $\Lambda_{i}^{\mu\nu}$-s.
Finally we will obtain
\begin{eqnarray}
\label{D4}
(U)^{\alpha}_{\alpha'}(U^{+})^{\beta'}_{\beta}&=&
(B_{2N}^{-1})^{\alpha}_{\gamma}(A_{1}^{+})^{\beta'}_{\delta'}
\left[{1 \over 2}(\sigma_1)^{\gamma}_{\delta} 
(\sigma_1)^{\delta'}_{\gamma'}-\right.\nonumber\\
&-&\left.{1 \over 2}(\sigma^{\mu} \sigma_1)^{\delta'}_{\gamma'} 
(\prod_{i=1}^{N} \Lambda_{2i-1} \Lambda_{2i})^{\mu\nu}
(\sigma^{\nu} \sigma_1)^{\gamma}_{\delta}\right]
(A_1)^{\gamma'}_{\alpha'}
(B_{2N}^{-1})^{+ \delta}_{\beta}.
\end{eqnarray}

Substituting this expression, together with the expressions 
for $B_{2N}^{-1}$ and $A_{1}$ (from (\ref{A30}) and 
(\ref{A29}) correspondingly) into the (\ref{RO}),
after some simple algebra for Landauer resistance $\rho$
we will have
\begin{eqnarray}
\label{RO2}
\rho&=&{1 \over 2}(\frac{\mu_1k_1}{\mu_2k_2})\left[-1+
(\Lambda^N)^{11} {1 \over 2}(\frac{\mu_1k_1}{\mu_2k_2}+
\frac{\mu_2k_2}{\mu_1k_1})+\right.\nonumber\\
&+&\left. i(\Lambda^N)^{12} {1 \over 2}(\frac{\mu_1k_1}{\mu_2k_2}-
\frac{\mu_2k_2}{\mu_1k_1})\right],
\end{eqnarray}
where $(\Lambda^N)^{11}$ (correspondingly $(\Lambda^N)^{12}$) is
the $11 (12)$ matrix element of the matrix $\Lambda =\Lambda_1 
\Lambda_2$, which is a product of $\Lambda$-s of the $I$ and $II$
slices.

The average over any type of random distributions of the 
parameters of the model can be calculated now exactly. We consider
random distribution of thicknesses of the slices, keeping
boundaries fixed $x_0 = 0, x_{2N} = L$. We see from the formula
(\ref{D2}) that $T_i$ depends only on the thickness of the slice
$x_i - x_{i-1}$. The only restriction we have is the condition, 
that
\begin{equation}
\label{B2}
\sum_{i=1}^{2N}\Delta x_i = L.
\end{equation}

Therefore, the average of the $\Lambda^N$, with the probability
distribution $g(y);\;\ (\int_0^{\infty} g(y)dy = 1$), defined 
in the following way
\begin{eqnarray}
\label{B3}
\langle \prod_{i=1}^{N}\Lambda_{2i-1}\Lambda_{2i}\rangle &=&
\int_{0}^{\infty}dy_{1}\dots dy_{2N} g(y_1)\dots g(y_{2N})
\delta(\sum_{j=1}^{2N}y_j - L) \prod_{i=1}^{N}
\Lambda_{2i-1}(y_{2i-1})\Lambda_{2i}(y_{2i})=\nonumber\\
&=& \int_{-\infty}^{\infty}dp e^{-ipL}\left(\langle\Lambda_{1}(p)
\rangle\langle\Lambda_{2}(p)\rangle\right)^N,
\end{eqnarray}
where
\begin{equation}
\label{B4}
\langle\Lambda_{1,2}(p)\rangle = \int_{0}^{\infty}
dy e^{ipy} g(y) \Lambda_{1,2}(y).
\end{equation}

The average Landauer resistance is equal now
\begin{eqnarray}
\label{B5}
\langle\rho \rangle&=&{1 \over 2}(\frac{\mu_1k_1}{\mu_2k_2})\left\{- 1+
{1 \over 2}(\frac{\mu_1k_1}{\mu_2k_2}+
\frac{\mu_2k_2}{\mu_1k_1})\int_{-\infty}^{\infty}dp e^{ipL}
\left[\left(\langle\Lambda_{1}(p)\rangle
\langle\Lambda_{2}(p)\rangle\right)^N 
\right]^{11}+\right.\nonumber\\
&+&\left.{1 \over 2}(\frac{\mu_1k_1}{\mu_2k_2}-
\frac{\mu_2k_2}{\mu_1k_1})
\int_{-\infty}^{\infty}dp e^{ipL}
\left[\left(\langle\Lambda_{1}(p)\rangle                    
\langle\Lambda_{2}(p)\rangle\right)^N 
\right]^{12}\right\}.
\end{eqnarray}
It is obvious, that in a case of homogeneous
media (two components of the superlattice are coinciding)
we restore the expression for the Landauer resistance
of electrons, obtained in \cite{EH}.

For large sample size $(N >> 1)$, as it was argued in 
\cite{La, Abrik}, the resistance should behave as $e^{N/\xi}$,
where $\xi$ is the correlation length. Excitations are
localized or not depends on the behaviour of $\xi$. If
at some frequencies correlation length becomes infinite,
we have delocalized state and the expression (\ref{B5})
shows, that the answer depends on average value of
$\Lambda^{\mu\nu}$. For further analyse let's consider simplest
case of the distribution, namely when there is equal
probability for slices to have a thickness up to $d_i$
(i=1,2).
\begin{equation}
\label{G}
g(y) = \left\{\begin{array}{l} \frac{1}{d_i},\;\;\;\;\;\; 0 < y <d_{i}\\
0,\;\;\;\;\;\;\; otherwise\end{array}\right..
\end{equation}
We have taken $Al_{x}Ga_{1-x}As$ and $GaAs$ as a components of the 
superlattice with the parameters \cite{24}
\begin{eqnarray}
\label{G1}
\mu_1&=& (3.25-0.09x)10^{11} dyn/cm^{2},\;\;\;
\mu_2 = 3.25 10^{11} dyn/cm^{2},\nonumber\\
\rho_1&=&(5.3176-1.6x) g/cm^3,\;\;\;\;\;\;\;\;\;\;\;\;
\rho_2 = 5.3176 g/cm^3,\nonumber\\
d_1&=&30\cdot (5.6532+0.0078x) A^{o},\;\;\;\;\;\;\;\;\;\;\;
d_2 = 10\cdot 5.6532 A^{o}
\end{eqnarray}
and consider waves, propagating in the perpendicular to
the layers direction $(\vec q =0)$.

For large enough $N$  the asymptotics of $\rho$, and
therefore the correlation length $\xi$, defined by the
closest to unity eigenvalues of $\langle\Lambda_1\rangle
\langle\Lambda_2\rangle$. If it is $\lambda$, then
\begin{equation}
\label{L}
\xi(\omega)\sim 1/\ln \lambda(\omega).
\end{equation}

Numerical calculations by use of Mathematica shows, that 
$\lambda(\omega=0)=1$, hence $\xi \rightarrow \infty$.

\setlength{\unitlength}{3947sp}%
\begingroup\makeatletter\ifx\SetFigFont\undefined%
\gdef\SetFigFont#1#2#3#4#5{%
  \reset@font\fontsize{#1}{#2pt}%
  \fontfamily{#3}\fontseries{#4}\fontshape{#5}%
  \selectfont}%
\fi\endgroup%
\begin{picture}(6099,4149)(289,-3598)
\thinlines
\put(4801,-361){\vector( 0, 1){900}}
\put(4801,-361){\line( 0,-1){150}}
\put(4801,-511){\line( 0,-1){150}}
\put(4801,-661){\line( 0,-1){150}}
\put(4801,-811){\line( 0,-1){150}}
\put(4801,-961){\line( 0,-1){150}}
\put(4801,-1111){\line( 0,-1){150}}
\put(4801,-1261){\line( 0,-1){150}}
\put(4801,-1411){\line( 0,-1){150}}
\put(4801,-1561){\line( 0,-1){150}}
\put(4801,-1711){\line( 0,-1){150}}
\put(4801,-1861){\line( 0,-1){150}}
\put(4801,-2011){\line( 0,-1){800}}
\put(451,-2686){\line( 2, 1){5850}}
\put(6001,-2161){\vector( 1, 0){375}}
\put(6001,-2161){\line(-1, 0){300}}
\put(5701,-2161){\line(-1, 0){300}}
\put(5401,-2161){\line(-1, 0){300}}
\put(5101,-2161){\line(-1, 0){600}}
\put(4501,-2161){\line(-1, 0){300}}
\put(4201,-2161){\line(-1, 0){300}}
\put(3901,-2161){\line(-1, 0){300}}
\put(3601,-2161){\line(-1, 0){300}}
\put(3301,-2161){\line(-1, 0){300}}
\put(3001,-2161){\line(-1, 0){300}}
\put(2701,-2161){\line(-1, 0){300}}
\put(2401,-2161){\line(-1, 0){300}}
\put(2101,-2161){\line(-1, 0){300}}
\put(1801,-2161){\line(-1, 0){600}}
\put(1201,-2161){\line(-1, 0){900}}
\put(1201,-2561){\shortstack{A}}
\put(1201,-2152){\shortstack{\circle{40}}}
\put(1151,-2061){\shortstack{-3}}
\put(2401,-2152){\shortstack{\circle{40}}}
\put(2351,-2361){\shortstack{-2}}
\put(3601,-2152){\shortstack{\circle{40}}}
\put(3551,-2361){\shortstack{-1}}
\put(6001,-2152){\shortstack{\circle{40}}}
\put(5951,-2361){\shortstack{1}}
\put(4801,-1552){\shortstack{\circle{40}}}
\put(4901,-1552){\shortstack{-20}}
\put(4801,-952){\shortstack{\circle{40}}}
\put(4901,-952){\shortstack{-18}}
\put(4801,-352){\shortstack{\circle{40}}}
\put(4901,-352){\shortstack{-16}}
\put(3901,152){\shortstack{$lnln\lambda(\omega$)}}
\put(6301,-2061){\shortstack{$ln\omega$}}
\put(2951,-3061){\shortstack{Fig.2}}
\end{picture}

This result is easy to understand, $\omega = 0$ means
that we have constant displacement $\vec u$, which
simply is the sift of the all sample. The correlation index
$\nu$ from $\xi \sim \omega^{-\nu}$, defined as a slop of
the plot of $lnln\lambda(\omega)$ versus $ln\omega$
is presented in Fig.2 and it appeared that $\nu$=2.
All other states are localized.
\section{Acknowledgement}
\indent
The authors acknowledge S.Badalyan and A.Khachatryan for
many valuable discussions. The work of D.G.S. was partially supported
by CRDF Grant-375100. 

\end{document}